\documentclass[aps,prb,twocolumn,amsmath,10pt]{revtex4-1}
\pdfoutput=1

\usepackage[ansinew]{inputenc}
\usepackage{graphicx}
\usepackage{textcomp}
\usepackage{hyperref}
\hypersetup{
pdftitle={A photon-photon quantum gate based on a single atom in an optical resonator},
pdfauthor={Bastian~Hacker, Stephan~Welte, Gerhard~Rempe, Stephan~Ritter}}

\newcommand{\ket}[1]{\ensuremath{\vert {#1}\rangle}} 
\newcommand{\up}{{\uparrow}} 
\newcommand{\down}{{\downarrow}} 
\newcommand{\micro}{\ensuremath{\textnormal\textmu}}
\newcommand{\unit}[1]{\ensuremath{\,\mathrm{#1}}}
\newcommand{\unnumsec}[1]{\refstepcounter{section}\section*{#1}}

\makeatletter
\renewcommand{\fnum@figure}{\textbf{Figure~\thefigure}}
\renewcommand{\@caption@fignum@sep}{ \textbf{\textbar}}
\@ifpackageloaded{hyperref}{\newcommand{\positionlabel}[1]{\hypertarget{#1}{}}}{\newcommand{\positionlabel}[1]{}}
\@ifpackageloaded{hyperref}{}{\newcommand{\hyperref}[2][]{#2}}
\@ifpackageloaded{hyperref}{\newcommand{\figref}[1]{\hyperlink{#1}{\ref*{#1}}}}{\newcommand{\figref}[1]{\ref{#1}}}
\makeatother

\begin{document}
\title{A photon-photon quantum gate based on a single atom in an optical resonator}

\author{Bastian~Hacker}
\thanks{These authors contributed equally to this work.}
\author{Stephan~Welte}
\thanks{These authors contributed equally to this work.}
\author{Gerhard~Rempe}
\author{Stephan~Ritter}
\email[To whom correspondence should be addressed. Email: ]{stephan.ritter@mpq.mpg.de}

\affiliation{Max-Planck-Institut f\"ur Quantenoptik, Hans-Kopfermann-Strasse 1, 85748 Garching, Germany}

\maketitle
\textbf{Two photons in free space pass each other undisturbed. This is ideal for the faithful transmission of information, but prohibits an interaction between the photons as required for a plethora of applications in optical quantum information processing \cite{Kok2010}. The long-standing challenge here is to realise a deterministic photon-photon gate. This requires an interaction so strong that the two photons can shift each others phase by $\boldsymbol{\pi}$. For polarisation qubits, this amounts to the conditional flipping of one photon's polarisation to an orthogonal state. So far, only probabilistic gates \cite{Knill2001} based on linear optics and photon detectors could be realised \cite{OBrien2003}, as ``no known or foreseen material has an optical nonlinearity strong enough to implement this conditional phase shift [\ldots]'' \cite{OBrien2007}. Meanwhile, tremendous progress in the development of quantum-nonlinear systems has opened up new possibilities for single-photon experiments \cite{Chang2014}. Platforms range from Rydberg blockade in atomic ensembles \cite{Gorshkov2011} to single-atom cavity quantum electrodynamics \cite{Reiserer2015}. Applications like single-photon switches \cite{Baur2014} and transistors \cite{Tiarks2014,Gorniaczyk2014}, two-photon gateways \cite{Kubanek2008}, nondestructive photon detectors \cite{Reiserer2013b}, photon routers \cite{Shomroni2014} and nonlinear phase shifters \cite{Turchette1995,Tiecke2014,Volz2014,Beck2015,Tiarks2016} have been demonstrated, but none of them with the ultimate information carriers, optical qubits in discriminable modes. Here we employ the strong light-matter coupling provided by a single atom in a high-finesse optical resonator to realise the Duan-Kimble protocol \cite{Duan2004} of a universal controlled phase flip (CPF, $\boldsymbol{\pi}$ phase shift) photon-photon quantum gate. We achieve an average gate fidelity of $\boldsymbol{\overline{F}=(76.2\pm3.6)\%}$ and specifically demonstrate the capability of conditional polarisation flipping as well as entanglement generation between independent input photons. Being the key quantum logic element, our gate could readily perform most of the hitherto existing two-photon operations. It also discloses avenues towards new quantum information processing applications where photons are essential, especially long-distance quantum communication and scalable quantum computing.}

The perhaps simplest idea to realise a photonic quantum gate is to overlap the two photons in a nonlinear medium. However, it has been argued that this cannot ensure full mutual information transfer between the qubits for locality and causality reasons \cite{Shapiro2006,Gea2010}. Instead, a viable strategy is to keep the two photons separate, change the medium with the first one, use this change to affect the second photon, and, finally, make the first photon interact with the medium again to ensure gate reciprocity. These three subsequent interactions enable full mutual information exchange between the two qubits, as required for a gate, even though the photons never meet directly.

Our experimental realisation of a CPF photon-photon gate builds on the proposal by Duan and Kimble \cite{Duan2004}. The medium is a single atom strongly coupled to a cavity and the interactions happen upon reflection of each photon off the atom-cavity system \cite{Reiserer2014}. While the proposal considers three reflections, we replace the second reflection of the first photon by a measurement of the atomic state and classical phase feedback on the first photon (analogous to a proposal \cite{Duan2005} where the roles of light and matter are interchanged). In practise, this allows us to achieve better fidelities, higher efficiencies and to use a simpler setup compared to that of the proposed scheme.

\begin{figure}
\centering
\positionlabel{fig:setup}
\includegraphics[width=\columnwidth]{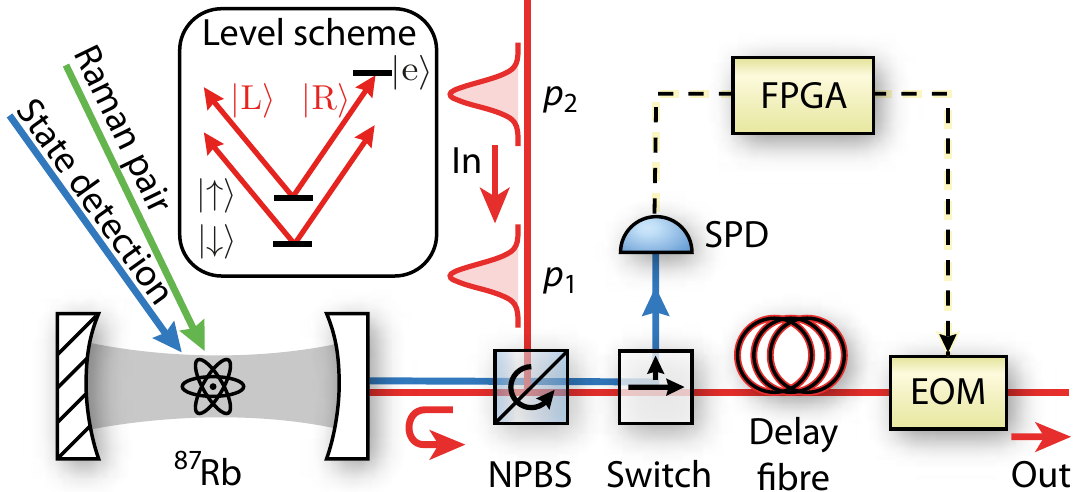}
\caption{\label{fig:setup}
\textbf{Schematic of our setup.} Qubit-carrying weak coherent photon pulses $p_1$ and $p_2$ enter in two separate spatio-temporal modes via a non-polarising 98.5\% transmitting beam splitter (NPBS) that acts effectively as a circulator. The photons are subsequently reflected from the cavity containing a single atom before a switch directs them into a delay fibre. While $p_1$ and $p_2$ are stored in the fibre, the state of the atom is read-out via fluorescence photons (blue arrows) that the switch directs towards a single-photon detector (SPD). A field programmable gate array (FPGA) applies a conditional phase feedback to $p_1$ via an electro-optical modulator (EOM). Eventually, the photons leave the gate setup towards polarisation analysers. The inset shows the atomic energy level scheme. The three depicted, relevant levels of $^{87}$Rb and the photon polarisations are defined in the main text. The photons and the empty cavity are on resonance with the atomic transition $\ket{\up}\leftrightarrow\ket{\text{e}}$.}
\end{figure}

\begin{figure}
\centering
\positionlabel{fig:scheme}
\includegraphics[width=\columnwidth]{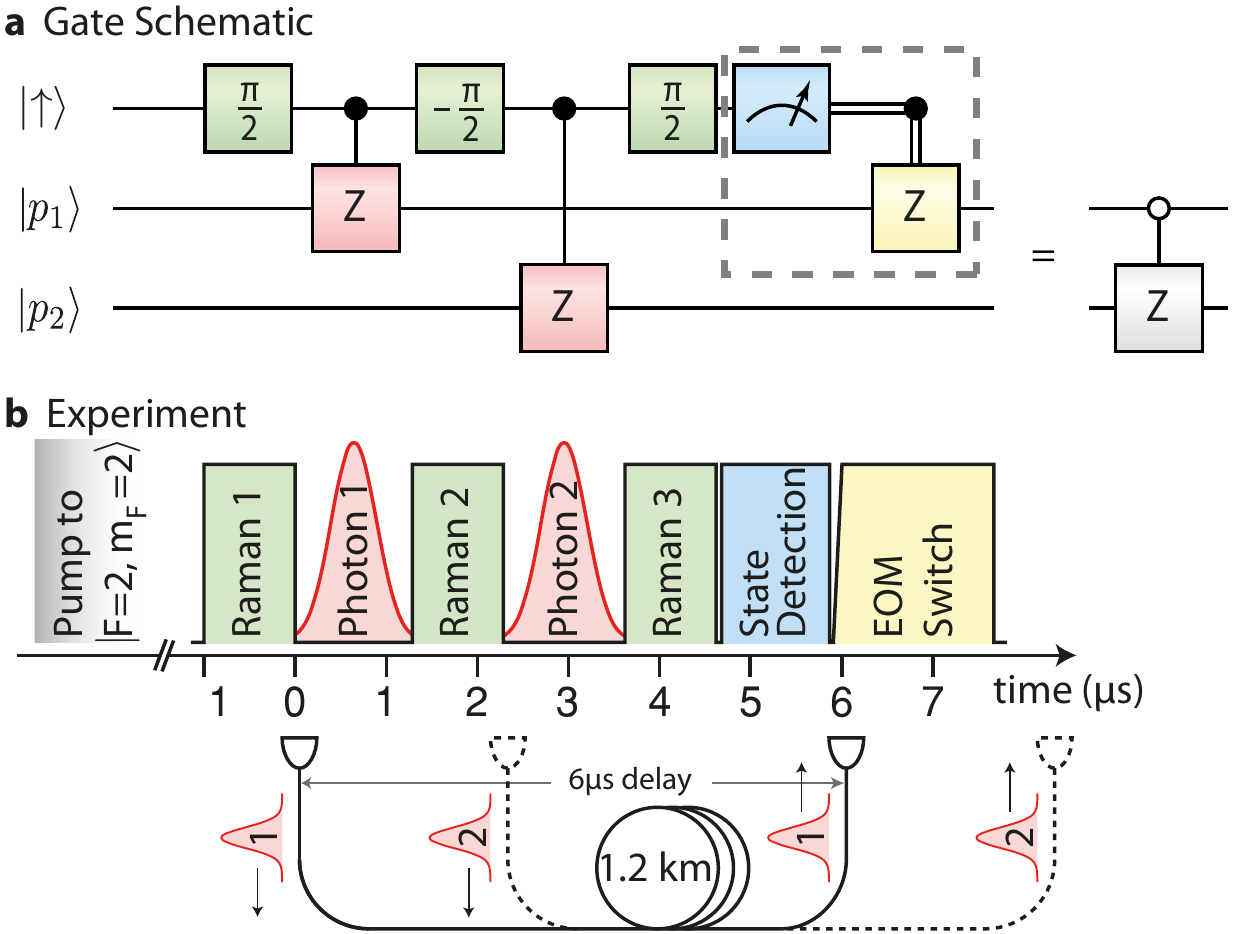}
\caption{\label{fig:scheme}
\textbf{The photon-photon gate mechanism.}
\textbf{a}, Quantum circuit diagram. The sequence of CZ gates between the atomic ancilla qubit and the gate photons interleaved with rotations on the atomic qubit acts as a pure CPF gate on the input photon state $\ket{p_1p_2}$. Note that the dashed box is equivalent to the reflection-based quantum CZ gate of the original proposal via the principles of deferred and implicit measurement. \textbf{b}, Pulse sequence showing the timing of the experimental steps of the gate protocol. A delay fibre of $1.2\unit{km}$ length is used to store the gate photons for $6\unit{\micro s}$.}
\end{figure}

We employ a single $^{87}$Rb atom trapped in a three-dimensional optical lattice \cite{Reiserer2013} at the centre of a one-sided optical high-finesse cavity \cite{Reiserer2013b} (Fig.\,\figref{fig:setup}). The measured cavity quantum electrodynamics parameters on the relevant transition $\ket{\up}=\ket{F{=}2,m_F{=}2}\leftrightarrow \ket{\text{e}}=\ket{F{=}3,m_F{=}3}$ of the D$_2$ line are $(g,\kappa,\gamma)=2\pi\,(7,2.5,3)$\,MHz. The atom akes the role of an ancilla qubit, implemented in the basis $\ket{\down}=\ket{F{=}1,m_F{=}1}$ and $\ket{\up}$, with the quantisation axis along the cavity axis. Both photonic qubits are individually encoded in the polarisation using the notation $\ket{\mathrm L}$ and $\ket{\mathrm R}$ for a left- and a right-handed photon, respectively. They are consecutively coupled into the cavity beam path via a non-polarising beam splitter (98.5\% transmission) which plays the role of a polarisation-independent circulator. The photons as well as the empty cavity are on resonance with the transition $\ket{\up}\leftrightarrow\ket{\text{e}}$ at $780\unit{nm}$. Only the atom in $\ket{\up}$ and the photon in $\ket{\mathrm R}$ are strongly coupled, because the $\ket{\down}\leftrightarrow\ket{\text{e}}$ transition is detuned by the ground-state hyperfine splitting of 6.8\,GHz, and the left-circularly polarised transition from $\ket{\up}\leftrightarrow \ket{F{=}3,m_F{=}1}$ is shifted out of resonance by a dynamical Stark shift induced by the laser that traps the atom. The strong light-matter coupling between $\ket{\up}$ and $\ket{\mathrm R}$ shifts the phase of a reflected photon by $\pi$ compared to the cases where the atom occupies $\ket{\down}$ or the photon is in $\ket{\text{L}}$. Thus, each reflection constitutes a bidirectional controlled-Z (CZ) interaction\cite{Reiserer2014} between the atomic and photonic qubit (red boxes in Fig.\,\figref{fig:scheme}a).

Figure \figref{fig:scheme}a depicts the experimental implementation of the photon-photon gate as a quantum circuit diagram. In short, the protocol starts with arbitrary photonic input qubits $\ket{p_1}$ and $\ket{p_2}$ and the atom optically pumped to $\ket{\up}$. After this initialisation, two consecutive atomic-qubit rotations combined with CZ atom-photon quantum gates are performed. The purpose of the rotations is to maximize the effect of the subsequent gates. Note that up to this point the first photon has the capability to act via the atom onto the second photon. To implement a back-action of the second photon onto the first one, the protocol ends with a measurement of the atomic qubit and feedback onto the first photon. This measurement has the additional advantage that it removes any possible entanglement of the atom with the photons, as required for an ancillary qubit. A longer and \hyperref[methods:composition]{detailed stepwise analysis} of the above protocol as well as the \hyperref[methods:rotations]{characterisation of the Raman lasers} used for the implementation of the atomic-state rotations can be found in the Methods.

To apply this scheme in practise, the qubits have to be stored and controlled in an appropriately timed sequence: After the first photon $p_1$ is reflected, it directly enters a $1.2\unit{km}$ delay fibre. The delay time of $6\unit{\micro s}$ is sufficient to allow for reflection of both photons from the cavity, two coherent spin rotations, and state detection on the atom (Fig.\,\figref{fig:scheme}b). The two photon wave packets are in independent spatio-temporal modes which can in principle be arbitrarily shaped. The only requirement is that the frequency spectrum should fall within the acceptance bandwidth of the cavity ($0.7\unit{MHz}$ for $\pm0.1\pi$ phase shift accuracy). We used Gaussian-like envelopes of $0.6\unit{\micro s}$ full width at half maximum (FWHM) within individual time windows of $1.3\unit{\micro s}$ width, such that the corresponding FWHM bandwidth of $0.7\unit{MHz}$ leads to an acceptable phase-shift spread.

After the last spin rotation, Purcell-enhanced fluorescence state detection of the atomic qubit is performed. This is achieved within $1.2\unit{\micro s}$ with a laser beam resonant with the $\ket{\up}\leftrightarrow\ket{\text{e}}$ transition and impinging perpendicular to the cavity axis (blue beam in Fig.\,\figref{fig:setup}). This yields zero fluorescence photons for $\ket{\down}$ and a near-Poissonian-distributed photon number  with an average of 4 for $\ket{\up}$, resulting in a discrimination fidelity of 96\%. The fluorescence light shares the same spatial mode as the gate photons and needs to be detected before the first photon leaves the delay fibre. Separation of the fluorescence light from the qubit photons is achieved with an efficient free-space acousto-optical deflector (AOD, labelled `Switch' in Fig.\,\figref{fig:setup}). Qubit photons pass the deactivated AOD straight towards the delay fibre, whereas state-detection photons are deflected into the first diffraction order directed at a single-photon detector (SPD). The corresponding detection events are evaluated in real time by a field programmable gate array (FPGA), which activates a $\pi$ phase shift on the $\ket{\mathrm R}$ component of the first gate photon if the atom was detected in $\ket{\up}$. No phase shift is applied if the atom was found in $\ket{\down}$. This conditional phase shift is performed by an electro-optical modulator (EOM) with a switching time of $0.1\unit{\micro s}$, which is ready when $p_1$ leaves the delay fibre and is reset before $p_2$ appears at the end of the fibre. The experiment runs at a rate of $500\unit{Hz}$, with each execution preceded by atom cooling, atomic state preparation via optical pumping and probing of cavity transmission to confirm success of the initialisation. All experiments with one detected qubit photon in each of the two temporal output modes are evaluated without further post-selection.

If both input photons are circularly polarised, the photon-photon gate appears as a CPF gate (see \hyperref[methods:composition]{Methods}) characterised by:
\begin{align*}
&\ket{\mathrm{RR}}\rightarrow\ket{\mathrm{RR}}
&&\ket{\mathrm{LR}}\rightarrow-\ket{\mathrm{LR}}\\
&\ket{\mathrm{RL}}\rightarrow\ket{\mathrm{RL}}
&&\ket{\mathrm{LL}}\rightarrow\ket{\mathrm{LL}}.
\end{align*}
As with any quantum gate, it can also be expressed in other bases. We define the linear polarisation bases as $\ket{\mathrm H}=\frac{1}{\sqrt{2}}(\ket{\mathrm R}{+}\ket{\mathrm L})$, $\ket{\mathrm V}=\frac{1}{\sqrt{2}}(\ket{\mathrm R}{-}\ket{\mathrm L})$, $\ket{\mathrm D}=\frac{1}{\sqrt{2}}(\ket{\mathrm R}{+}i\ket{\mathrm L})$, and $\ket{\mathrm A}=\frac{-1}{\sqrt{2}}(i\ket{\mathrm R}{+}\ket{\mathrm L})$, respectively. With one of the photons being circularly and the other one linearly polarised, the gate will act as a controlled-NOT (CNOT) gate with the circular qubit being the control and the linear one being the target qubit. When both photons enter in linear polarisation states, the gate will turn the two separable inputs into a maximally entangled state.

We characterised the gate by applying it to various pairs of separable input-qubit combinations and by measuring the average outcome from a large set of repeated trials. The input were two independent weak coherent pulses each impinging with an average photon number of $\overline{n}=0.17$ onto the cavity. The choice of $\overline{n}$ is a compromise between measurement time and measured gate fidelity. While lowering $\overline{n}$ reduces the data rate because of the high probability of zero-photon events in either of the two photon modes, increasing $\overline{n}$ raises the multi-photon probability per pulse thereby deteriorating the measured gate fidelity.

\begin{figure}
\centering
\positionlabel{fig:truthtable}
\includegraphics[width=0.96\columnwidth]{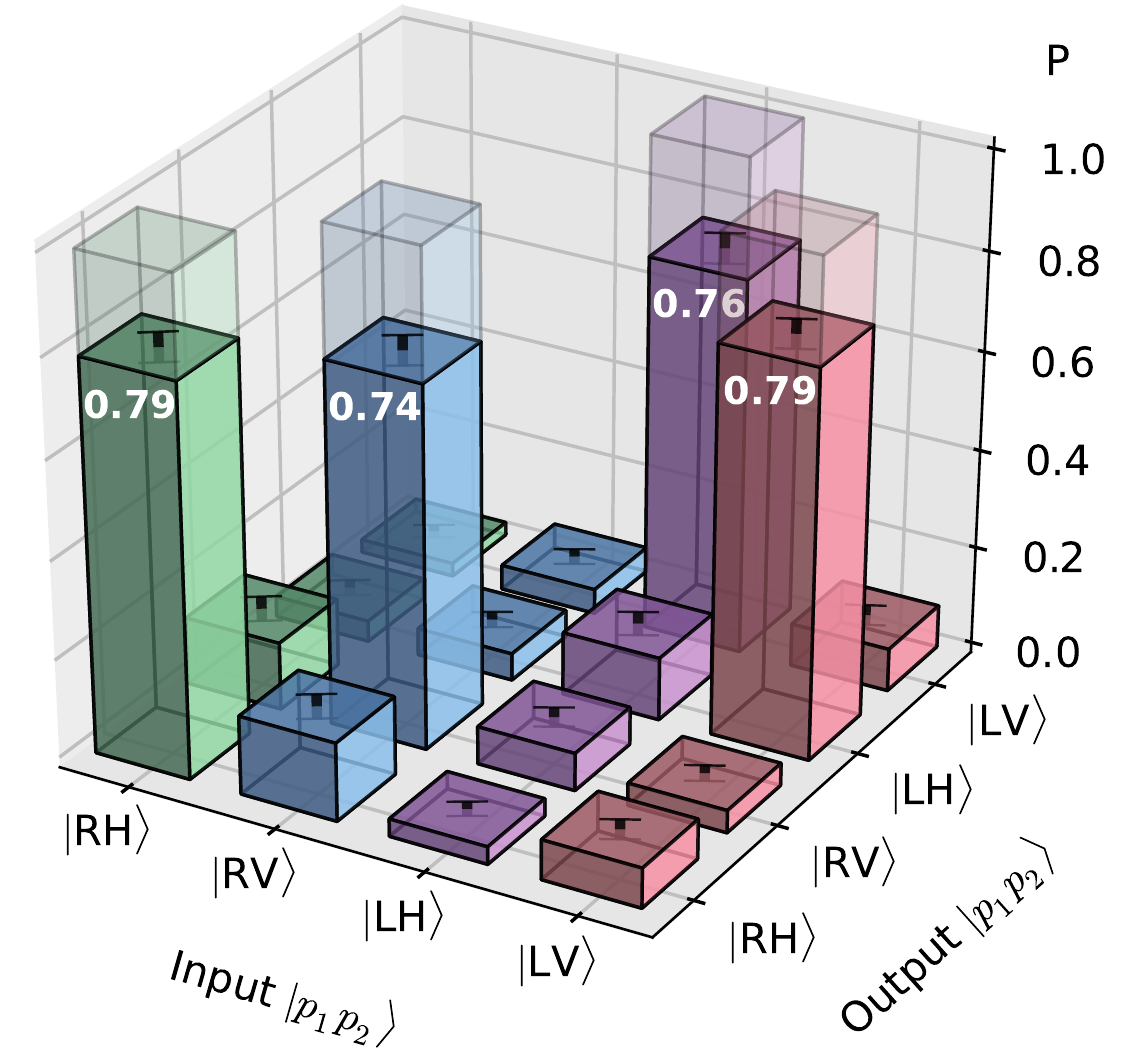}
\caption{\label{fig:truthtable}
\textbf{Truth table of the CNOT photon-photon gate.} The gate flips the linear polarisation of the target photon $p_2$ if the control photon $p_1$ is in the state $\ket{\mathrm L}$, while it leaves the target qubit unchanged if the control photon is in $\ket{\mathrm R}$. The vertical axis gives the probability to measure a certain output state given the designated input state. The truth table for an ideal CNOT gate is indicated by the four light transparent bars with $\text{P}=1$. The black T-shaped bars represent statistical errors on each entry (rms 2.2\%), computed via linear error propagation assuming independent photon statistics.}
\end{figure}

First, we processed the four different input states of a CNOT basis, i.e.\ all combinations of photon $p_1$ in the circular basis and $p_2$ in a linear basis, and analysed them in the corresponding measurement bases. The resulting truth table is depicted in Fig.\,\figref{fig:truthtable} and shows an overlap with the case of an ideal CNOT gate of $F_\textnormal{CNOT}=(76.9\pm1.5)\%$.

\begin{figure}
\centering
\positionlabel{fig:densitymatrix}
\includegraphics[width=0.96\columnwidth]{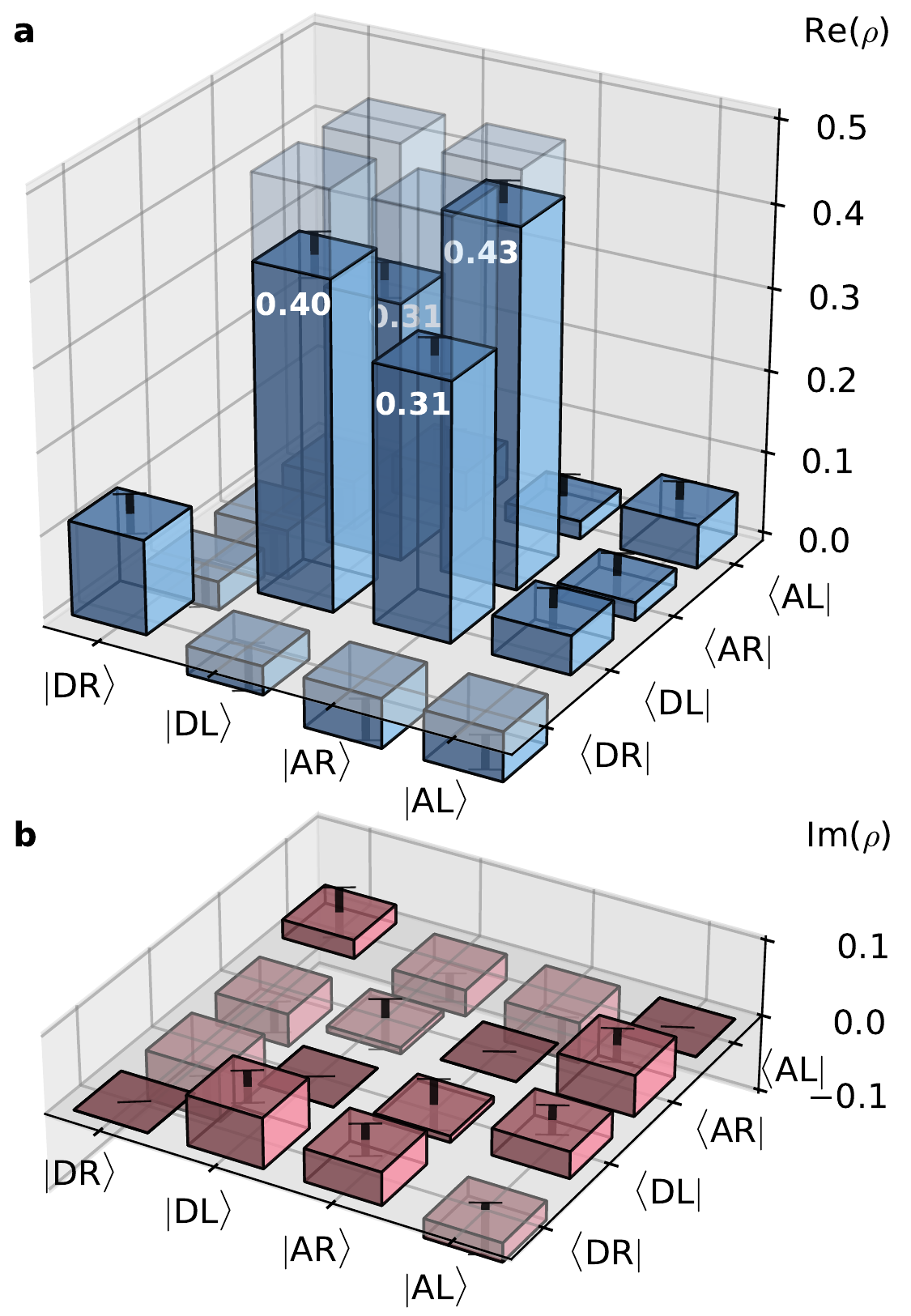}
\caption{\label{fig:densitymatrix}
\textbf{Reconstructed density matrix of the entangled two-photon state created by the gate from the separable input state $\boldsymbol{\ket{\mathrm{DD}}}$.} Depicted are the real~(\textbf{a}) and imaginary~(\textbf{b}) part of the elements of the density matrix. The light transparent bars indicate the ideal density matrix for $\ket{\Psi^+}$ in the chosen basis. Statistical errors on each entry (rms 2.4\%) are drawn as black T-shaped bars.}
\end{figure}

A decisive property of a quantum gate that distinguishes it from its classical counterpart is its capability to generate entanglement. For both input photons in the linear polarisation state $\ket{\mathrm D}$, the gate ideally creates the maximally entangled Bell state $\ket{\Psi^+}=\frac{1}{\sqrt{2}}(\ket{\mathrm{DL}}+\ket{\mathrm{AR}})$. We reconstructed the output of the gate for the input state $\ket{\mathrm{DD}}$ from 1378 detected photon pairs via linear inversion and obtained the density matrix $\rho$ depicted in Fig.\,\figref{fig:densitymatrix}. It has a fidelity $F_{\Psi^+}=\langle\Psi^+|\rho|\Psi^+\rangle=(72.9\pm 2.8)\%$ with the ideal Bell state (unbiased linear estimate). The generation of this entangled state from a separable input state directly sets a non-tight bound for the entangling capability (smallest eigenvalue of the partially transposed density matrix)\cite{Poyatos1997} of our gate, $\mathcal{C}\leq-0.242\pm0.028$, which is $-0.5$ for the ideal CPF gate and where a negative $\mathcal{C}$ denotes that the gate is entangling. We remark that the total data set can be separated into two subsets of equal size corresponding to the outcome of the atomic state detection being $\ket{\down}$ or $\ket{\up}$. The respective fidelities are $F^{\down}_{\Psi^+}=(74.4\pm 3.9)\%$ and $F^{\up}_{\Psi^+}=(71.5\pm 4.2)\%$, i.e.\ the gate works comparably well in both cases.

As an overall measure of the gate performance we determined the average gate fidelity $\overline{F}$, which is equal to the average fidelity of $6\times6$ output states generated from the input states on all canonical polarisation axes (H, V, D, A, R, L) with the theoretically expected ideal outcomes \cite{Bagan2003}. All 36 state fidelities were estimated linearly and bias-free with randomised tomographically complete basis settings. Although we collected only insignificant statistics of 80 detected photon pairs on each of the output states, their combination gives a meaningful measure of $\overline{F}=(76.2\pm3.6)\%$. The deviation from unity is well understood for our system and results from technical imperfections which we discuss below.

The efficiency of the presented gate, which is the combined transmission probability for two photons, is unity for the ideal scheme, but gets reduced by several experimental imperfections. It is polarisation-independent because all optical elements including the cavity have near-equal losses for all polarisations. The two main loss channels are the long delay fibre (transmission $T=40.4\%$) and the limited cavity reflectivity ($R=67\%$). The latter results from the cavity not being perfectly single-sided and a finite cooperativity of $C=3.3$. All other optical elements have a combined transmission of $81\%$, dominated by the fibre-coupling efficiency and absorption of the AOD switch. This yields a total experimental gate efficiency of $(22\%)^2=4.8\%$. Despite the transmission losses, characteristic for all photonic devices, the protocol itself is deterministic. The largest potential improvement is offered by eliminating the fibre-induced losses, for instance by a free-space delay line, a delay cavity or an efficient optical quantum memory.

We have modelled all known sources of error (see \hyperref[methods:simulation]{Methods}) to reproduce the deviation of the experimental gate fidelity from unity. Here, we quote the reductions in fidelity that each individual effect would introduce to an otherwise perfect gate. The largest contribution stems from using weak coherent pulses to characterise the gate and is therefore not intrinsic to the performance of the gate itself. First, there is a significant probability of having two photons in one qubit mode if it is populated, resulting in a phase flip of $2\pi$ instead of $\pi$, causing an overall reduction of the gate fidelity by 12\%. Second, the probability to have both qubit modes populated is small, such that detector dark counts contribute 2\% error. The measured gate fidelity could therefore be greatly improved by employing a true single-photon source \cite{Reiserer2015}.

The relatively short delay introduced by the optical fibre restricts the temporal windows for the photon pulses and atomic state detection. The resulting bandwidth of the photons reduces the gate fidelity by 6\%. The obvious solution is to choose a longer delay. Further errors can be attributed to the characteristics of the optical cavity (5\%), the state of the atom (6\%), and other optical elements (2\%). The cavity has a polarisation-eigenmode splitting of $420\unit{kHz}$ that could be eliminated by mirror selection \cite{Uphoff2015}. Neither the resonance frequency of the cavity nor the spatial overlap between its mode and the fibre mode are perfectly controlled (see \hyperref[methods:modematching]{Methods}). The latter could be improved with additional or better optical elements. Fidelity reductions associated with the state of the atom are due to imperfect state preparation, manipulation and detection, and decoherence. Improvements are expected from the application of cavity-enhanced state detection to herald successful state preparation, Raman sideband cooling to eliminate variations in the Stark shift of the atom, and composite pulses to optimise the state rotations. The limited precision of polarisation settings and polarisation drifts inside the delay fibre are the main contribution from other optical elements. The latter could be improved using active stabilization. The wealth of realistic suggestions for improvement given above shows that progress towards even higher fidelities is certainly feasible for the presented gate implementation.

The photon-photon gate as first demonstrated here follows a deterministic protocol and could therefore be a scalable building block for new photon-processing tasks such as those required by quantum repeaters \cite{Briegel1998}, for the generation of photonic cluster states \cite{Raussendorf2001} or quantum computers \cite{Ladd2010}. The gate's ability to entangle independent photons could be a resource for quantum communication. Moreover, our gate could serve as the central processing unit of an all-optical quantum computer, envisioned to processes pairs of photonic qubits that are individually stored in and retrieved from an in principle arbitrarily large quantum cache. Such cache would consist of an addressable array of quantum memories, individually connected to the gate via optical fibres. Eventually, such architecture might even be implemented with photonic waveguides on a chip.

We thank Norbert Kalb, Andreas Neuzner, Andreas Reiserer and Manuel Uphoff for fruitful discussions and support throughout the experiment.
This work was supported by the European Union (Collaborative Pro\-ject SIQS) and by the Bundesministerium f\"ur Bildung und Forschung via IKT 2020 (Q.com-Q) and by the Deutsche Forschungsgemeinschaft via the excellence cluster Nanosystems Initiative Munich (NIM). S.\,W. was supported by the doctorate program Exploring Quantum Matter (ExQM).

\vfill
This work:
\newenvironment{extrabibitem}{\list{}{\leftmargin=1em\rightmargin=0em}\item[]}{\endlist}
\begin{extrabibitem}
\small
Hacker, B., Welte, S., Rempe, G. \& Ritter, S.
A photon-photon quantum gate based on a single atom in an optical resonator.
\href{http://dx.doi.org/10.1038/nature18592}{\emph{Nature} \textbf{536}, 193--196} (2016).
\end{extrabibitem}

\newpage

\renewcommand{\figurename}{\textbf{Extended Data Figure}}
\renewcommand{\thefigure}{\arabic{figure}}
\makeatletter
     \@addtoreset{figure}{section}
\makeatother

\unnumsec{Methods}

\noindent\textbf{\phantomsection\label{methods:composition}Composition of the photon-photon CPF gate.}
The action of the quantum circuit diagram depicted in Fig.\,\figref{fig:scheme}a can be computed in the eight-dimensional Hilbert space spanned by the atomic ancilla qubit and the two photonic qubits. The atomic single-qubit rotations by $\pi/2$ and $-\pi/2$ are described by the operators $\frac{1}{\sqrt{2}}\left(\begin{smallmatrix}1&-1\\1&1\end{smallmatrix}\right)$ and $\frac{1}{\sqrt{2}}\left(\begin{smallmatrix}1&1\\-1&1\end{smallmatrix}\right)$, respectively, in the basis $\{\ket{\up},\ket{\down}\}$. The atom-photon CZ-gate is described by $U_{ap}=\mathrm{diag}(-1,1,1,1)$ in the basis $\{\ket{\up\mathrm R},\ket{\up\mathrm L},\ket{\down\mathrm R},\ket{\down\mathrm L}\}$. As indicated in Fig.\,\figref{fig:scheme}a, the atom is initially prepared in $\ket{\up}$. Any input state of the two photonic qubits, including entangled states, can be written as
\[
\ket{p_1p_2} = c_{\mathrm{RR}}\ket{\mathrm{RR}} + c_{\mathrm{RL}}\ket{\mathrm{RL}} + c_{\mathrm{LR}}\ket{\mathrm{LR}} + c_{\mathrm{LL}}\ket{\mathrm{LL}},
\]
defined by the four complex numbers $c_{\mathrm{RR}}$, $c_{\mathrm{RL}}$, $c_{\mathrm{LR}}$ and $c_{\mathrm{LL}}$. Henceforth, we will use the compact notation $\ket{rr}:=c_{\mathrm{RR}}\ket{\mathrm{RR}}$, $\ket{rl}:=c_{\mathrm{RL}}\ket{\mathrm{RL}}$, $\ket{lr}:=c_{\mathrm{LR}}\ket{\mathrm{LR}}$, and $\ket{ll}:=c_{\mathrm{LL}}\ket{\mathrm{LL}}$.
Therefore, any photon-photon gate operation starts in the collective initial state
\[
\ket{\up}(\ket{rr}+\ket{rl}+\ket{lr}+\ket{ll}).
\]
The first $\pi/2$ rotation brings the atom into a superposition
\[
\textstyle\frac{1}{\sqrt{2}}(\ket{\up}+\ket{\down})\ (\ket{rr}+\ket{rl}+\ket{lr}+\ket{ll}),
\]
followed by a CZ-interaction between the atom and the first photon, which flips the sign of all states with the atom in $\ket{\up}$ and the first photon in $\ket{\mathrm R}$:
\[
\textstyle\frac{1}{\sqrt{2}}\bigl((-\ket{\up}+\ket{\down})(\ket{rr}+\ket{rl})+(\ket{\up}+\ket{\down})(\ket{lr}+\ket{ll})\bigr).
\]
Subsequent rotation of the atom by $-\pi/2$ creates the state
\[
\ket{\down}(\ket{rr}+\ket{rl}) + \ket{\up}(\ket{lr}+\ket{ll}).
\]
Reflection of the second photon flips the sign of all states with the atom in $\ket{\up}$ and the second photon in $\ket{\mathrm R}$:
\[
\ket{\down}(\ket{rr}+\ket{rl}) + \ket{\up}(-\ket{lr}+\ket{ll}).
\]
The final rotation of the atom by $\pi/2$ yields
\[
\textstyle\frac{1}{\sqrt{2}}\bigl((-\ket{\up}+\ket{\down})(\ket{rr}+\ket{rl})\; + \;
(\ket{\up}+\ket{\down})(-\ket{lr}+\ket{ll})\bigr).
\]
At this point the state of the atom is measured. There are two equally probable outcomes projecting the two-photon state accordingly:
\begin{align*}
\ket{\up}{:}\quad& -\ket{rr}-\ket{rl}-\ket{lr}+\ket{ll},\\
\ket{\down}{:}\quad& +\ket{rr}+\ket{rl}-\ket{lr}+\ket{ll}.
\end{align*}
Following detection of the atom $\ket{\up}$, an additional $\pi$ phase is imprinted on the $\ket{\mathrm R}$-part of the first photon, i.e.\ a sign flip on $\ket{rr}$ and $\ket{rl}$, whereas the photonic state is left unaltered upon detection of $\ket{\down}$. Thereby, the final photonic state becomes
\[
\ket{rr}+\ket{rl}-\ket{lr}+\ket{ll},
\]
independent of the outcome of the atomic state detection. It differs from the input state by a minus sign on $\ket{lr}$ only. Hence, the total circuit acts as a pure photonic CPF gate:
\begin{align*}
&\ket{\mathrm{RR}}\rightarrow\ket{\mathrm{RR}}
&&\ket{\mathrm{LR}}\rightarrow-\ket{\mathrm{LR}}\\
&\ket{\mathrm{RL}}\rightarrow\ket{\mathrm{RL}}
&&\ket{\mathrm{LL}}\rightarrow\ket{\mathrm{LL}}.
\end{align*}

\vspace{\baselineskip}\noindent\textbf{\phantomsection\label{methods:rotations}Calibration of atomic single-qubit rotations.}
In order to calibrate relevant experimental parameters, we employ a Ramsey-like sequence of three subsequent rotation pulses. The pulses are exactly timed as in the gate sequence (see Fig.\,\figref{fig:scheme}), but the two photon pulses interleaved between the Raman pulses are turned off.

\begin{figure}[b]
\centering
\positionlabel{fig:threepulse}
\includegraphics[width=\columnwidth]{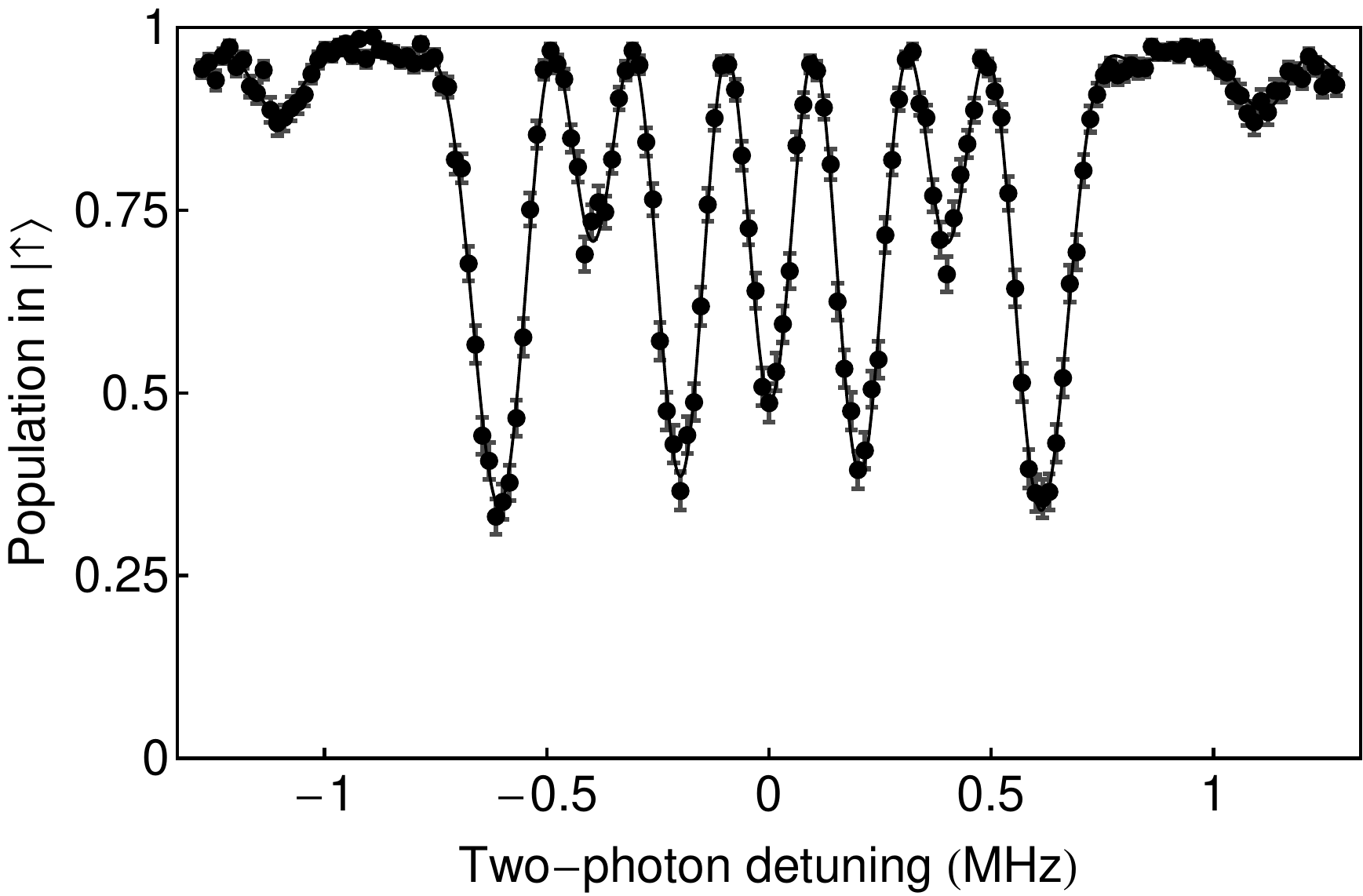}
\caption{\label{fig:threepulse}
\textbf{Ramsey-like spectrum to calibrate the atomic state rotations.}
After initialisation of the atom in $\ket{\up}$, we perform the same sequence of three Raman pulses as in the gate protocol. The final population in $\ket{\up}$ is determined as a function of the two-photon detuning of the employed Raman pair with respect to the frequency difference between the two atomic qubit states. The solid dots are measured data with statistical error bars. The solid line is the fit of a theoretical model based on the sequence of rotations. It yields results for the Rabi frequency of the atomic spin rotation, an offset of the two-photon detuning, as e.g.\ induced by ambient magnetic fields, and the light shift imposed by the Raman laser pair, all with $\pm 3\unit{kHz}$ precision.}
\end{figure}

Initially, the atom is prepared in $\ket{\up}$. The Raman pair is red-detuned by $131\unit{GHz}$ from the D$_1$ line of $^{87}$Rb. Employing an acousto-optic modulator, we scan one of the Raman lasers over $2.5\unit{MHz}$ while the frequency of the other is fixed. Thus, we effectively scan the two-photon detuning. Extended Data Fig.\,\figref{fig:threepulse} shows a spectrum depicting the population in $\ket{\up}$ as a function of the two-photon detuning. Ideally, the gate experiments are performed on two-photon resonance. In this case, the second pulse compensates the first and the third one brings the atom into the superposition state $(\ket{\up}+\ket{\down})/\sqrt{2}$, such that $50\%$ population in $\ket{\up}$ are obtained.

To determine the experimental parameters that guarantee this situation, a theoretical model is fitted to the spectrum. It allows us to simultaneously access several mutually dependent fit parameters useful to calibrate the frequency as well as the intensity of our Raman beams. The fit reveals the Rabi frequency for the transition between $\ket{\down}$ and $\ket{\up}$, which we tune to $250\unit{kHz}$ to obtain $\pi/2$ pulses in $1\unit{\micro s}$. The two-photon detuning is also extractable from the fit and we find a light shift of $40\unit{kHz}$ due to the Raman lasers. To compensate for it, we choose different two-photon detunings when the pulses are on and off, such that two-photon resonance is guaranteed during the entire sequence.

\vspace{\baselineskip}\noindent\textbf{\phantomsection\label{methods:modematching}Transverse optical mode matching.}
Good overlap between the transverse mode profiles of the incoming wave packet and the optical cavity is essential for the performance of the gate. To this end, the qubit-carrying photon pulses are taken from a single-mode fibre with its mode matched to the cavity. In a characterisation measurement we determined that 92\% of probe light emanating from the cavity is coupled into this input fibre. Therefore, 8\% of the impinging light may arrive in an orthogonal mode that does not interact with the atom-cavity system. Light in this mode deteriorates the fidelity of the gate if it is collected at the output. This problem is overcome, because the delay fibre also acts as a filter for the transverse mode profile after the cavity. The mode overlap between cavity and delay fibre is 84\%, partially suffering from mode distortion by the AOD used for path switching. From an analysis of cavity reflection spectra we can estimate the amount of light that did not interact with the cavity but is still coupled from the input fibre into the delay fibre. It is below 1\% of the gate output, such that the resulting reduction of the gate fidelity is also well below 1\%.

A small misalignment, e.g.\ due to slow temperature drifts, reduces the positive filtering effect described above. Therefore, optimal mode matching is essential to maintain maximum gate fidelity. In the experiment, reflection spectra of the empty cavity were constantly monitored and, whenever necessary, data taking was interrupted to reestablish optimal mode overlap.

\vspace{\baselineskip}\noindent\textbf{\phantomsection\label{methods:simulation}Simulation of imperfections.}
In order to understand the imperfections encountered in the experiment, we have set up a model of both photonic qubits and the atomic ancilla qubit in terms of their three-particle density matrix $\rho$. Under ideal conditions, the density matrix transforms via sequential unitary transformations $U$ as $\rho\rightarrow U\rho U^\dagger$, and known error sources can be introduced at each specific step. Finally, the fidelity of $\rho$ with the desired target state is calculated for comparison with the experimental value.

In this scenario, an unnoticed, incorrect preparation of the atom creates an incoherent admixture of the wrong initial state. Errors in the atomic state detection lead to an exchange of the photonic submatrices corresponding to each atomic state. Detector dark counts are modeled as an admixture of a fully mixed state and decoherence effects are taken into account as reductions in off-diagonal elements of $\rho$. Cases where photons do not enter the cavity because of geometric mode mismatch are included with a phase shift of zero, and the case of an undetected additional photon in one of the weak pulses is incorporated with a phase shift of $2\pi$, i.e.\ twice the ideal value. Interestingly, most deteriorations of the atom-photon interaction, like fluctuations of the atomic, cavity and photon frequencies, all condense into a variation, $\Delta\varphi=\pm0.15\pi$, of the conditional phase shift. Considering this together with the polarisation rotation $R_p(\xi)$ a photon experiences due to the residual cavity birefringence by an angle of $\xi=0.06\pi$ in case of $\ket{\down}$, the ideal atom-photon CZ-gate $U_{ap}=\mathrm{diag}(-1,1,1,1)$ in the basis
$\{\ket{\up\mathrm R},\ket{\up\mathrm L},\ket{\down\mathrm R},\ket{\down\mathrm L}\}$ must be replaced by:
\[
U_{ap} = \left[\begin{array}{cccc}
e^{i(\pi+\Delta\varphi)}&0&0&0\\
0&1&0&0\\
\begin{array}{c}0\\0\end{array}&\begin{array}{c}0\\0\end{array}&\multicolumn{2}{c}{R_p(\xi)}
\end{array}\right]
\]
Random fluctuations in some of the parameters enter our model by integrating the resulting density matrix over the assumed Gaussian distribution function.

\clearpage

\end{document}